%% file: main.tex
\documentclass[conference]{IEEEtran}
\IEEEoverridecommandlockouts
\usepackage{cite}
\usepackage{amsmath,amssymb,amsfonts}
\usepackage{bm}
\usepackage{algorithmic}
\usepackage{graphicx}
\usepackage{textcomp}
\usepackage{xcolor}
\usepackage{makecell}
\usepackage{multicol}
\usepackage{multirow}
\usepackage[labelsep=period]{caption}
\usepackage{siunitx}
\usepackage{subcaption}
\usepackage{url}

\def\BibTeX{{\rm B\kern-.05em{\sc i\kern-.025em b}\kern-.08em
    T\kern-.1667em\lower.7ex\hbox{E}\kern-.125emX}}

\title{Synthetic Speech Classification: IEEE Signal Processing Cup 2022 challenge}

\author{\IEEEauthorblockN{Mahieyin Rahmun$^1$, Rafat Hasan Khan$^1$, Tanjim Taharat Aurpa$^1$, Sadia Khan$^1$, Zulker Nayeen Nahiyan$^1$,}
\IEEEauthorblockN{Mir Sayad Bin Almas$^1$, Rakibul Hasan Rajib$^1$, Syeda Sakira Hassan$^2$
}
\IEEEauthorblockA{\textsuperscript{1}\textit{Independent University, Bangladesh}, \textsuperscript{2}\textit{Aalto University, Finland}}
}

\input{macros}

\begin{document}

\maketitle

\begin{abstract}
The aim of this project is to implement and design a robust synthetic speech classifier for the IEEE Signal Processing Cup 2022 challenge. Here, we learn a synthetic speech attribution model using the speech generated from various text-to-speech (TTS) algorithms as well as unknown TTS algorithms. We experiment with both the classical machine learning methods such as support vector machine, Gaussian mixture model, and deep learning based methods such as ResNet, VGG16, and two shallow end-to-end networks. We observe that deep learning based methods with raw data demonstrate the best performance.
\end{abstract}


\section{Introduction}
Human-like speech generation from text is no longer a challenge. With the recent advancement in text-to-speech (TTS) algorithms, powerful models, for example, Tacotron~\cite{wang2017tacotron, skerry2018towards}, WaveNet~\cite{vandenoord16ssw}, and DeepVoice~\cite{arik2017deep} can produce almost natural human voice. This raises the possibility of manipulating someone's voice data~\cite{fletcher2018deepfakes}, which is an alarming issue and can have severe negative impact on the society. Therefore, it is important than ever before to detect such maliciously altered voices. In recent years, many spoof detection challenges had been organized~\cite{wu2015asvspoof, wu2017asvspoof, Todisco2019ASVspoof2F}. Interested readers are referred to~\cite{hua2021towards} and references therein. However, in this challenge, we will explore solutions to a synthetic speech attribution problem, i.e., finding out the type of algorithms generate synthetic speech. 

The rest of the paper is structured as follows. In Section~\ref{section:data}, we discuss the properties of the dataset provided by the SPCUP2022 competition and data preparation technique. In Section~\ref{section:methods}, we give an overview of both classical and deep learning approaches that we employ for experimenting on the datasets. In Section~\ref{section:feature_analysis}, we explore how well the features are separated for classification. Subsequently, in Section~\ref{section:experimental_results}, we discuss the experimental results, and in Section~\ref{section:team_contibutions}, we describe the contributions of the team members. Finally, in Section~\ref{section:discussion_and_conclusion}, we draw concluding remarks.

\section{Data}
\label{section:data}

In this section, we discuss the training and evaluation data of Part 1 and 2 of the competition as well as the noisy dataset that is used in Part 2 of the competition. We also provide an overview on dataset preparation.
\subsection{Training data}
\subsubsection{Part 1}
\label{subsection:dataset_part_1}
This dataset is composed of 5000 noise-free audio samples. There are a total of 5 classes, labeled 0 through 4. Each class represents a certain speech synthesizer algorithm that was used to generate the audio samples. Table~\ref{tab:properties_data} illustrates the properties of the audio files from dataset Part 1. However, the identity of the algorithm is hidden. Each class consists of 1000 audio samples in the dataset. However, there is an uneven distribution of male and female voices within the dataset, with some classes having little to no male voices and vice-versa. Audio samples vary in length.

\begin{table}[tbp]
    \centering
    \caption{Properties of audio samples from dataset Part 1.}
      \begin{tabular}{ |c|c|c|c|c| } 
         \hline	
         Label & \thead{No.\\ of samples} & \thead{Total\\ duration (s)} & \thead{Mean\\ duration (s)} & \thead{Std.\\ deviation of\\ duration (s)} \\
         \hline
         0 & 1000 & 8255.93 & 8.26 &  2.75 \\ 
         \hline
         1 & 1000 &  6426.55 & 6.43 & 2.08	 \\
         \hline
          2 & 1000 & 6359.18	& 6.36	 & 	2.12 \\ 
         \hline
          3 & 1000 & 8136.84 & 8.14 & 	2.56 \\
          \hline
          4 & 1000 & 5615.91 & 	5.62 & 1.91  \\
         \hline
    \end{tabular}
    \label{tab:properties_data}
\end{table}

\subsubsection{Part 2}
This dataset is composed of 1000 noise-free audio samples, all of which belongs to a single class labeled as ``5''. This particular label implies that the audio samples were generated using multiple speech synthesizers not among those described in the previous section. The total duration in this case is 6791.28 seconds, mean duration is 6.79 seconds and standard deviation of duration is 2.22 seconds.

\subsection{Evaluation data}
This dataset also has two parts. In Part 1, we have 9000 noise-free audio samples with no label information although the samples are generated using audio synthesizing algorithms that are either known or unknown. The average total duration of the dataset is 61539.27 seconds, mean duration is 6.84 seconds and standard deviation of duration is 2.43 seconds. Similarly in Part 2, there are 9000 samples but they have been augmented using different augmentations such as noise injection, adding reverberations and MP3 compression. It is completely unknown as to which tracks have been augmented and how. The labels, average and total duration as well as the standard deviation of duration are similar to Part 1.

\subsection{Noisy data}
The second part of the competition emphasizes on synthetic speech attribution in the presence of noisy data. To facilitate this, MATLAB scripts are provided by the competition organizer which can perform the three augmentation techniques on audio samples discussed in the previous section. These augmentation are parametric and the exact values of the parameters during evaluation are not guaranteed to be fixed.

\subsection{Dataset preparation}
Since the evaluation dataset has no label information, in order to facilitate the initial training, we adopt stratified, non-overlapping splitting strategy of the provided training data into training, validation and testing sets. When no augmentation is applied, there are 4320 training, 480 validation and 1200 testing samples. However, using augmentation increases the amount of samples three-fold, resulting in 17280, 1920 and 4800 training, validation and testing samples respectively.

\section{Methods}
\label{section:methods}
In this section, we describe in brief the methodologies that we use for synthetic speech attribution. We explore both classical machine learning approaches and deep learning approaches.

\subsection{Classical machine learning models}

\subsubsection{Support vector machines (SVM)}
Given a dataset $D = \{\mathbf{d}_i, y_i\}$, where $\mathbf{d}_i \in \mathbb{R}^n$ is an audio sample belonging to class $y_i \in  \{0, \ldots, C\}, i = 1{\ldotp \ldotp}N$ and $C\in \mathcal{N}$. We are interested in building a model such that $\mathbf{y} = f(\psi(\mathbf{d}))$, where $\psi(.): \mathbb{R}^n \mapsto \mathbb{R}^m$. Here, $\psi(.)$ transforms $\mathbf{d}_i$ into a feature vector $\mathbf{x}_i \in \mathbb{R}^m$. To solve the multi-class classification problem, we need \mbox{$(C - 1)$} 2-class SVM~\cite{boser1992training} classifiers in one-versus-all setting. That is, the feature $\mathbf{x}_i$ belongs to class $C_i$ or the rest (not $C_i$) can be defined by the 2-class SVM classifier as $f(\mathbf{x}) = \operatorname{sign}(\mathbf{w}^{T}\mathbf{x} + b)$, where $\mathbf{w}$ is the weight vector, $b$ is the intercept term, and $\operatorname{sign}(\cdot)$ is the sign function. $f(\mathbf{x})$ takes the value of $+1$ if the sample belongs to class $C_i$ and $-1$ otherwise. The final decision of a sample belonging to class is taken by majority voting among all the $C-1$ classifiers.

\subsubsection{Gaussian mixture model (GMM)}
In audio signal classification, we can consider that an audio sample belongs to a class parameterized by some distribution such as Gaussian distribution. This motivates us to represent the features $\mathbf{x}$ of the sample as $\mathbf{x} \sim \mathcal{N}(\mathbf{\mu}_i, \mathbf{\Sigma}_i)$, where $\mathbf{\mu}_i$ and $\mathbf{\Sigma}_i$ are the mean and covariance of class $y_i$. Since the joint probability distribution of a particular class may be multi-modal, i.e.\ the samples of the class are distributed into multiple clusters, we make a mixture model composed of $C$ multivariate Gaussian distributions, each with mixture probability $\pi_i$ such that $ p(\mathbf{x} \mid \mathbf{\mu}, \mathbf{\Sigma}, \mathbf{\pi}) = \sum_{i=1}^C \pi_i\,p(\mathbf{x} \mid \mathbf{\mu}_i, \mathbf{\Sigma}_i) $. There are several approaches to learn this distribution by computing the maximum-likelihood estimates. For instance, one can use iterative optimization or expectation-maximization techniques~\cite{bishop}.

\subsection{Deep learning models}
\subsubsection{Baseline}
For establishing our baseline, we consider a shallow and end-to-end time-domain synthetic speech detection net (TSSDNet) proposed by~\cite{hua2021towards}, which is inspired from ResNet~\cite{he2016deep} and Inception-style~\cite{szegedy2015going} architectures. Henceforth, we refer to the ResNet-style TSSDNet architecture as Res-TSSDDNet and Inception-style architecture as Inc-TSSDNet. Both use 1D convolutions and batch-normalization, with the Inc-TSSDNet implementing dilated convolution~\cite{yu2015multi} unlike~\cite{szegedy2015going}, in order to benefit from increased receptive field of the model. We use the same architecture except the final layer, which we modify to accommodate the larger number of class labels, i.e. 5 without  ``unknown'' label and 6 with ``unknown'' label, in contrast to the original work. 

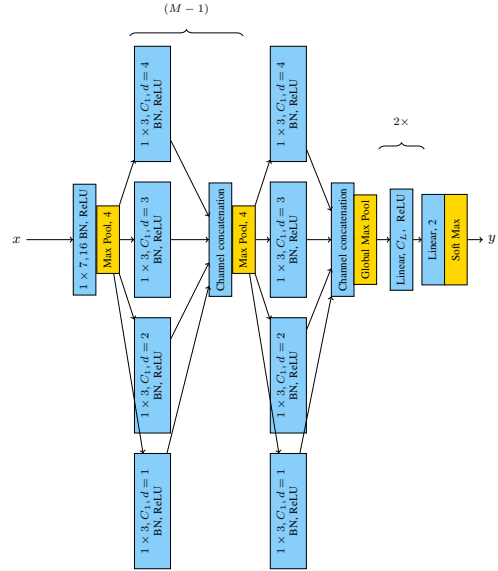
\begin{figure}[tbp]
    \centering
    \resizebox{0.75\columnwidth}{!}{\input{images/architectures/tssdnet/model_inc_tssdnet}}
    \caption{The Inception-style TSSDNet architecture (adapted from~\cite{hua2021towards}).}
    \label{fig:tssd_net}
\end{figure}

Both models accept raw audio signal as inputs, which we keep to be exactly 6 seconds in length during training and testing by utilizing either repetition (for shorter length waveform or trimming (for waveform with longer length than 6 seconds). The architecture of Inc-TSSDNet is illustrated in Figure~\ref{fig:tssd_net}.

\subsubsection{Convolutional deep learning methods}
For exploring more possibilities, we consider three convolutional neural networks: ResNet34, ResNet18 and VGG16 (see~\cite{He2016DeepRL} for more discussion). All the methods uses a stack of convolutional  layers containing filters of size 3x3 followed by fully connected layers~\cite{DBLP:journals/corr/SimonyanZ14a}. We incorporate batch normalization with VGG16 due to the poor performance of the standard VGG16. To build speech synthesis classifiers using these algorithms requires the raw data to be transformed to features, which can be done by using feature extraction methods, such as mel-frequency cepstral coefficients (MFCCs)~\cite{zheng2001comparison}.

\section{Feature Analysis}
\label{section:feature_analysis}
\begin{figure}[tpb]
     \centering
     \begin{subfigure}[b]{0.2\textwidth}
         \centering
         \includegraphics[width=\linewidth, height=0.9\linewidth]{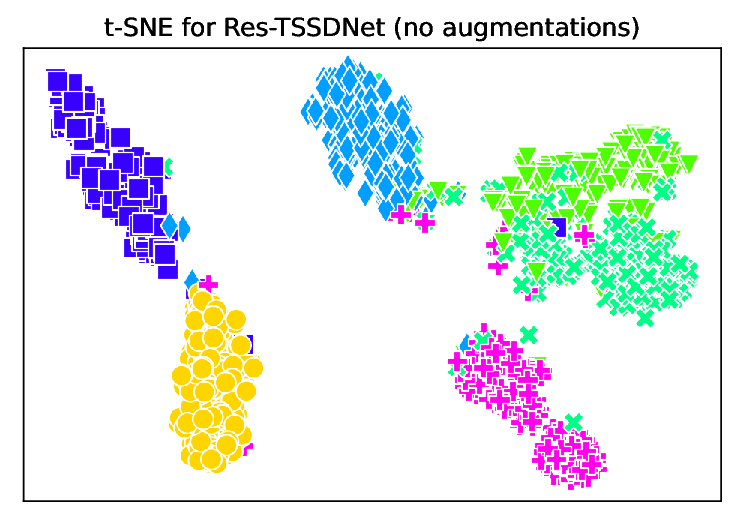}
         \caption{}
         \label{fig:feature_res_tssd_no_aug}
     \end{subfigure}
     \begin{subfigure}[b]{0.2\textwidth}
         \centering
         \includegraphics[width=\linewidth, height=0.9\linewidth]{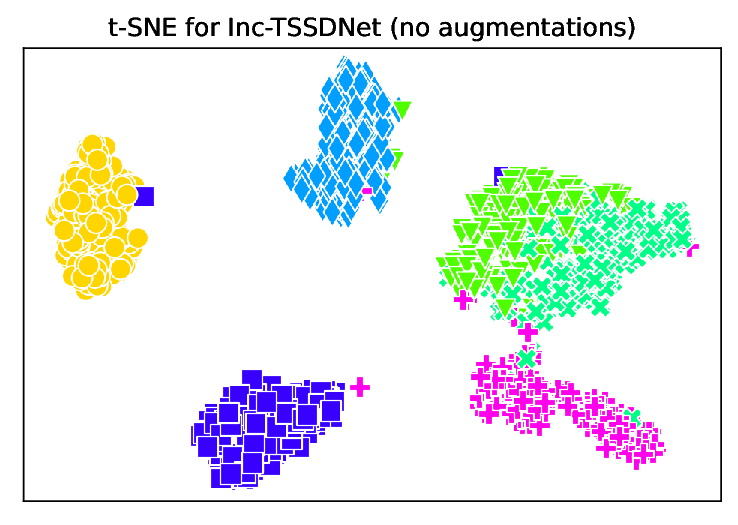}
         \caption{}
         \label{fig:feature_inc_tssd_no_aug}
     \end{subfigure}
     \begin{subfigure}[b]{0.2\textwidth}
         \centering
         \includegraphics[width=\linewidth, height=0.9\linewidth]{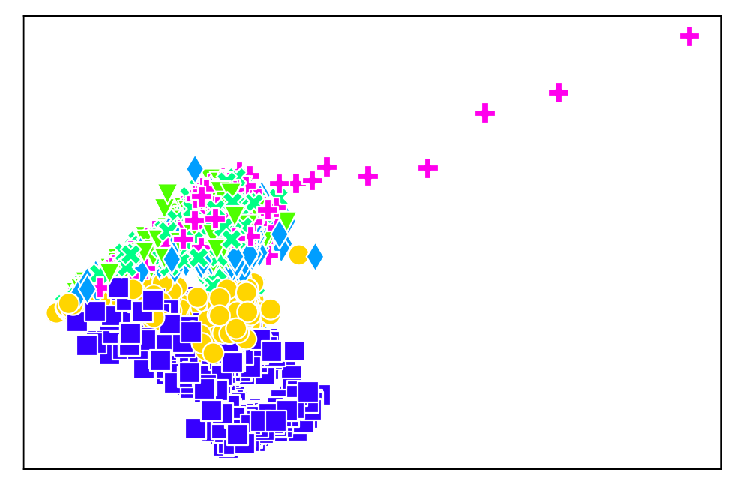}
         \caption{}
         \label{fig:feature_mfcc_no_aug}
     \end{subfigure}
     \begin{subfigure}[b]{0.2\textwidth}
         \centering
         \includegraphics[width=\linewidth, height=0.9\linewidth]{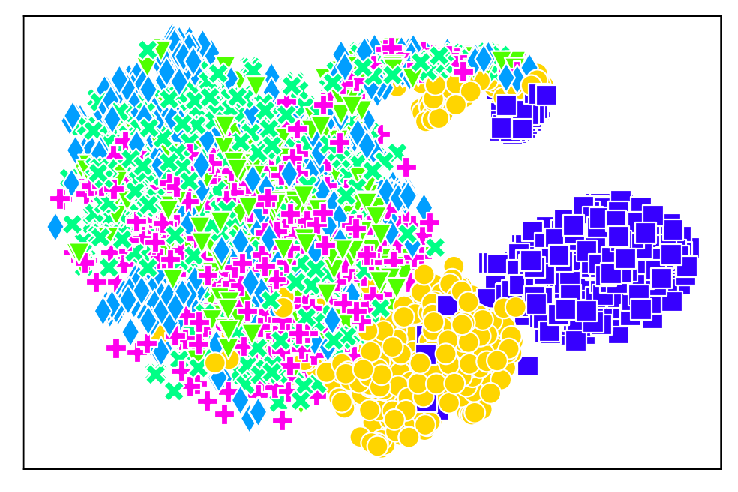}
         \caption{}
         \label{fig:feature_vgg16_aug}
     \end{subfigure}
     \begin{subfigure}[b]{0.2\textwidth}
         \centering
         \includegraphics[width=\linewidth, height=0.9\linewidth]{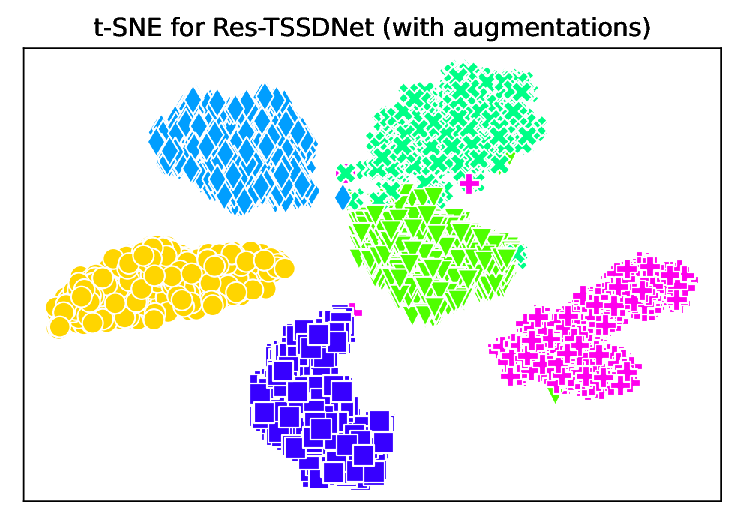}
         \caption{}
         \label{fig:feature_res_tssd_aug}
     \end{subfigure}
     \begin{subfigure}[b]{0.2\textwidth}
         \centering
         \includegraphics[width=\linewidth, height=0.9\linewidth]{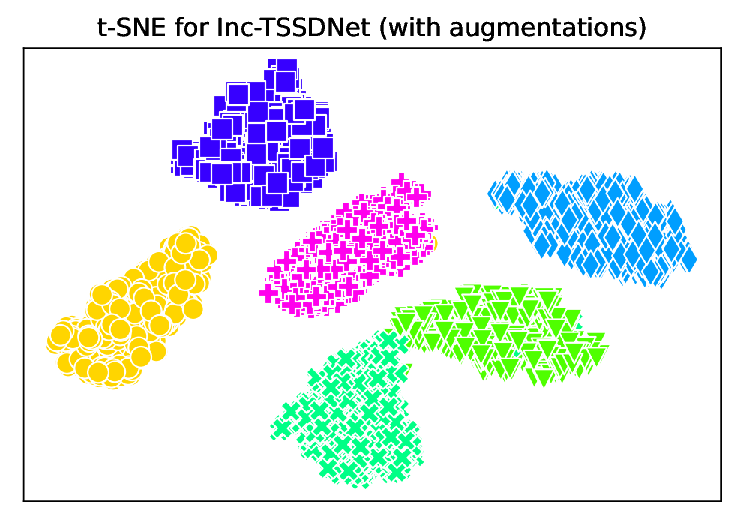}
         \caption{}
         \label{fig:feature_inc_tssd_aug}
     \end{subfigure}
    \caption{t-SNE plots of the feature embeddings for (a) Res-TSSDNet, no augmentations, (b) Inc-TSSDNet, no augmentations, (c) MFCC, with augmentation (d) VGG16, with augmentations, (e) Res-TSSDNet, with augmentations, and (f) Inc-TSSDNet, with augmentations. The colors blue, yellow, light blue, green, light green and pink represent class labels $\{0, 1, 2, 3, 4, 5\}$, respectively.}
    \label{fig:feature_embeddings}
\end{figure}

Fig.~\ref{fig:feature_embeddings} shows the t-distributed stochastic neighbor embedding (t-SNE)~\cite{van2008visualizing} plots of the feature embeddings for the methods discussed above. For the Res-TSSDNet and Inc-TSSDNet architectures, we take the output of the final linear layer just before the classification layer. These are 32-dimensional vectors. By contrast, mel-spectogram cepstral coefficients (MFCC) features were computed from raw waveform directly. In both cases, principal component analysis (PCA)~\cite{wold1987principal} was applied for dimensionality reduction, and subsequently the t-SNE plots were computed. It can be seen that Inc-TSSDNet achieves the maximum inter-class separation while preserving intra-class compactness regardless of whether augmented data is used. In contrast, MFCC fails to provide enough discriminative features for reliable classification.

\section{Experimental Results}
\label{section:experimental_results}
In this section, we describe the experimentation setup and report results for the architectures we have discussed in the previous section.

\subsection{Experimental setup and hyperparameters}
All the models were trained on a server machine running a 16-core Intel\textsuperscript{\textregistered} Xeon\textsuperscript{\textregistered} CPU E5-2640 v3 @ 2.60GHz, 8 Nvidia Tesla K80 GPUs with 12 GB virtual memory. For implementing the classical machine learning models, we use the scikit-learn~\cite{scikit-learn} library, while for implementing the deep learning based models, we use PyTorch with PyTorch Lightning~\cite{FalconPyTorchLightning2019} library. The implementations are available at GitHub\footnote{\url{https://github.com/AGenCyLab/SPCUP2022}}. We train all the models except GMM with 200 epochs with a batch size of 128 and an initial learning rate of $10^{-3}$. We use exponential learning rate scheduler with a learning rate decay factor $\gamma = 0.95$ per epoch. For TSSDNet type models, we used raw data as features, where as, with other models, we used different features listed in Table~\ref{tab:dl_accuracy}.

\subsection{Speech synthesis classifiers}
\begin{figure*}[tpb]
     \centering
     \begin{subfigure}[b]{0.3\textwidth}
         \centering
         \includegraphics[width=0.7\linewidth, height=0.7\linewidth]{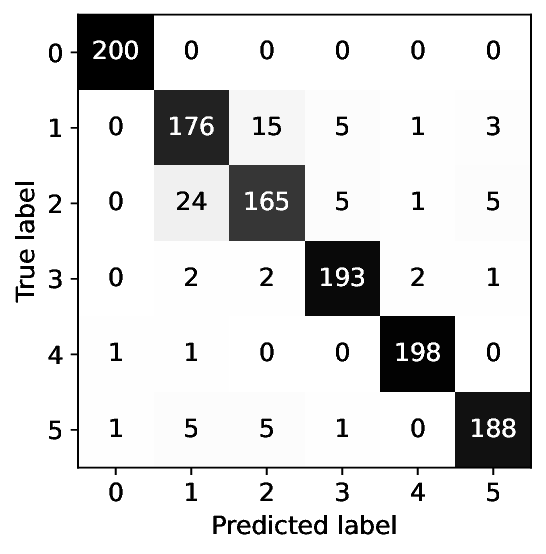}
         \caption{}
         \label{fig:cnf_matrix_a}
     \end{subfigure}
     \begin{subfigure}[b]{0.3\textwidth}
         \centering
         \includegraphics[width=0.7\linewidth, height=0.7\linewidth]{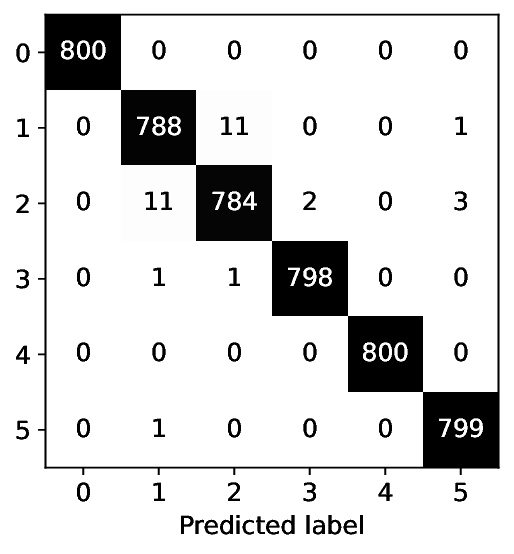}
         \caption{}
         \label{fig:cnf_matrix_b}
     \end{subfigure}
     \begin{subfigure}[b]{0.3\textwidth}
         \centering
         \includegraphics[width=0.7\linewidth, height=0.7\linewidth]{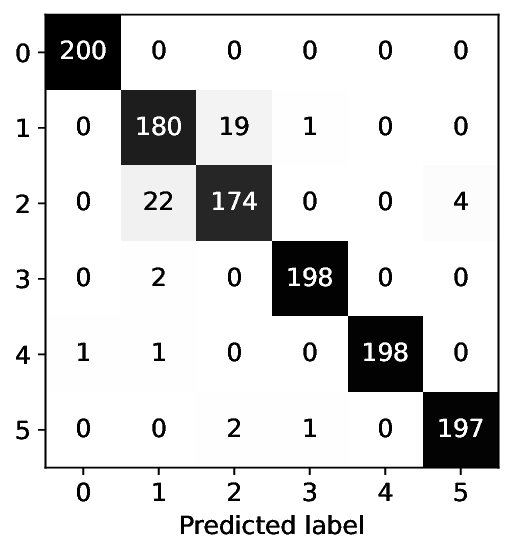}
         \caption{}
         \label{fig:cnf_matrix_c}
     \end{subfigure}
     \hfill
     \begin{subfigure}[b]{0.3\textwidth}
         \centering
         \includegraphics[width=0.7\linewidth, height=0.7\linewidth]{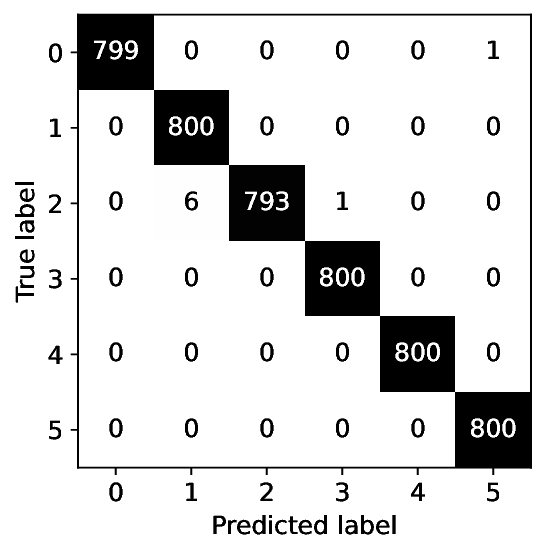}
         \caption{}
         \label{fig:cnf_matrix_d}
     \end{subfigure}
     \begin{subfigure}[b]{0.3\textwidth}
         \centering
         \includegraphics[width=0.7\linewidth, height=0.7\linewidth]{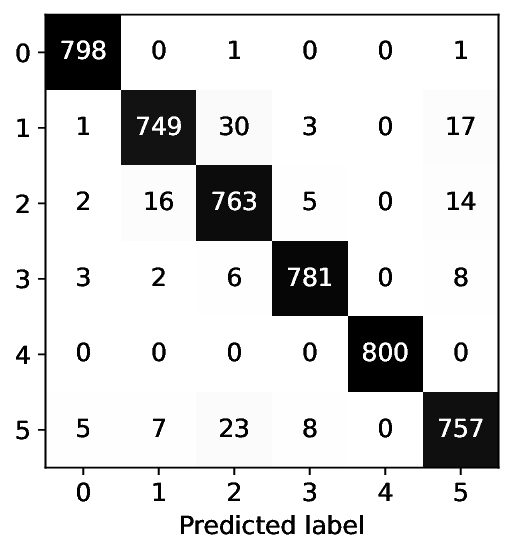}
         \caption{}
         \label{fig:cnf_matrix_e}
     \end{subfigure}
     \begin{subfigure}[b]{0.3\textwidth}
         \centering
         \includegraphics[width=0.7\linewidth, height=0.7\linewidth]{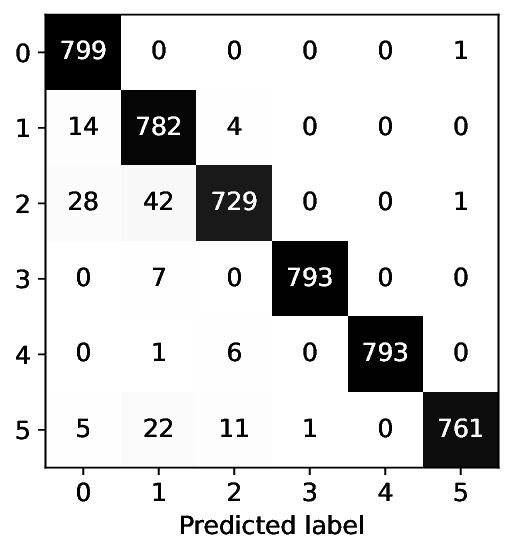}
         \caption{}
         \label{fig:cnf_matrix_f}
     \end{subfigure}
        \caption{Confusion matrices for (a) Res-TSSDNet, no augmented data, (b) Res-TSSDNet, augmented data, (c) Inc-TSSDNet, no augmented data, (d) Inc-TSSDNet, augmented data, (e) ResNet34, augmented data, and (f) SVM, augmented data.}
        \label{fig:cnf_matrices}
\end{figure*}

Figure \ref{fig:cnf_matrices} shows the confusion matrices of various classifiers that we described in Section~\ref{section:methods} on samples belonging to the held-out test set. It can be noted that after using augmented data, TSSDNet type models perform really well during testing. As Figure~\ref{fig:feature_embeddings} shows, this is possible due to the discriminative features learnt by the networks. Table~\ref{tab:dl_accuracy} summarizes the performance of the models.

\begin{table}[tbp]
\caption{Comparison of different deep learning methods. The accuracies reported are on the held-out testing set of the provided training data as discussed in Section~\ref{section:data}. The features are mel-spectrogram (MS), mel-frequency cepstral coefficients (MFCC), and raw waveform (RW).}
\begin{center}
\begin{tabular}{|c|c|c|c|}
\hline
Methods & Features & \multicolumn{2}{c|}{Accuracy}\\
\cline{3-4}
 & & Unaugmented data & Augmented data \\
\hline
ResNet34 & MS & 0.95 & 0.97 \\
\hline
ResNet18 & MS & \textbf{0.98} & 0.97 \\
\hline
VGG16 & MS & 0.70 & 0.68 \\
\hline
Inc-TSSDNet & RW & 0.96 & \textbf{1.00} \\
\hline
Res-TSSDNet & RW & 0.93 & 0.99 \\
\hline
SVM & MFCC & 0.86 & 0.97 \\
\hline
GMM & MFCC & 0.25 & 0.39 \\
\hline
\end{tabular}
\label{tab:dl_accuracy}
\end{center}
\end{table}

\section{Team members and contributions}
\label{section:team_contibutions}
Mahieyin Rahmun contributed in planning the experiment, implementing deep learning based models, and writing. Tanjim Taharat Aurpa  contributed in data and feature analysis as well as writing. Rafat Hasan Khan contributed in implementing deep learning based models and writing. Sadia Khan and Zulker Nayeen Nahiyan contributed in writing. Mir Sayad Bin Almas and Rakibul Hasan Rajib contributed in reviewing. Sakira Hassan contributed in reviewing the code and the report.

\section{Discussion and Conclusion}
\label{section:discussion_and_conclusion}
From Table~\ref{tab:dl_accuracy}, we observe that Inc-TSSDNet performs the best on augmented data with raw audio, while ResNet18 performs best on unaugmented data with mel-spectrogram features. Both Inc-TSSDNet and Res-TSSDNet see improvements in their performances with augmented data. We argue that both the larger number of training samples along with the numerous variations of the same audio contribute towards helping the models learn a robust feature space required for better classification. This is also observed when SVM classifier is trained with MFCC features. VGG16 has the worst performance among all, since it only considers mel-spectrogram features and does not have all required information to distinguish the classes. However, if we use a denser network such as ResNet34 (accuracy values are 0.95 and 0.97 for unaugmented and augmented data, respectively), the performance of the classifier improves. Based on our observations, we have decided to put forward Inc-TSSDNet as the detector of choice for synthetic speech attribution. According to Codalab\footnote{\url{https://codalab.lisn.upsaclay.fr}}, on a subset of the evaluation set part 1, the model achieves an accuracy of 0.84, while scoring 0.91 on evaluation set part 2, which is the highest among all the other models presented in this paper.

\section*{Acknowledgement}
Thanks to AGenCy Lab\footnote{\url{https://agencylab.github.io/}} for the computational utilities.

\bibliographystyle{IEEEtran}
\bibliography{reference}
\end{document}

%% file: macros.tex
\usepackage{tikz}

\definecolor{fc}{HTML}{1E90FF}
\definecolor{h}{HTML}{FFD700}
\definecolor{bias}{HTML}{87CEFA}
\definecolor{noise}{HTML}{8B008B}
\definecolor{conv}{HTML}{FFA500}
\definecolor{pool}{HTML}{B22222}
\definecolor{up}{HTML}{B22222}
\definecolor{view}{HTML}{FFFFFF}
\definecolor{bn}{HTML}{FFD700}
\tikzset{fc/.style={black,draw=black,fill=fc,rectangle,minimum height=1cm}}
\tikzset{h/.style={black,draw=black,fill=h,rectangle,minimum height=0.5cm}}
\tikzset{bias/.style={black,draw=black,fill=bias,rectangle,minimum height=0.5cm}}
\tikzset{noise/.style={black,draw=black,fill=noise,rectangle,minimum height=1cm}}
\tikzset{conv/.style={black,draw=black,fill=conv,rectangle,minimum height=1cm}}
\tikzset{pool/.style={black,draw=black,fill=pool,rectangle,minimum height=1cm}}
\tikzset{up/.style={black,draw=black,fill=up,rectangle,minimum height=1cm}}
\tikzset{view/.style={black,draw=black,fill=view,rectangle,minimum height=1cm}}
\tikzset{bn/.style={black,draw=black,fill=bn,rectangle,minimum height=1cm}}

\usetikzlibrary{decorations.pathreplacing,calligraphy, angles, quotes}

\newrobustcmd*{\mysquare}[1]{\tikz{\filldraw[draw=#1,fill=#1] (0,0) rectangle (0.2cm,0.2cm);}}

\newrobustcmd*{\mycircle}[1]{\tikz{\filldraw[draw=#1,fill=#1] (0,0) circle [radius=0.1cm];}}

\newrobustcmd*{\mytriangle}[1]{\tikz{\filldraw[draw=#1,fill=#1] (0,0) -- (0.2cm,0) -- (0.1cm,0.2cm);}}

%% file: images/architectures/tssdnet/model_inc_tssdnet.tex



\begin{tikzpicture}
    \node (x) at (1,0) {\small$x$};
    \node[bias, rotate=90] (r1) at (2.50,0) {\scriptsize{$1\times7, 16$  {BN, ReLU}} };
    \node[h,rotate=90] (h1) at (3.02,0) {\scriptsize{Max Pool, 4}};
    
    \node[bias, rotate=90, minimum width=2cm] (r21) at (4, -6) {\scriptsize\begin{tabular}{c}
           $1\times3, C_1, d=1$ \\$\text{BN, ReLU} $
    \end{tabular}};
    \node[bias, rotate=90, minimum width=2cm] (r22) at (4, -3) {\scriptsize\begin{tabular}{c}
           $1\times3, C_1, d=2$ \\$\text{BN, ReLU} $
    \end{tabular}};
    \node[bias, rotate=90, minimum width=2cm] (r23) at (4, 0) {\scriptsize\begin{tabular}{c}
           $1\times3, C_1, d=3$ \\$\text{BN, ReLU} $
    \end{tabular}};
    \node[bias, rotate=90, minimum width=2cm] (r24) at (4, 3) {\scriptsize\begin{tabular}{c}
           $1\times3, C_1, d=4$ \\$\text{BN, ReLU} $
    \end{tabular}};
    
    \node[bias, rotate=90] (r3) at (5.5,0) {\scriptsize{Channel concatenation} };
    \node[h,rotate=90,minimum width=1cm] (h2) at (6.02,0) {\scriptsize{Max Pool, 4}};

    \node[bias, rotate=90, minimum width=2cm] (r41) at (7, -6) {\scriptsize\begin{tabular}{c}
           $1\times3, C_1, d=1$ \\$\text{BN, ReLU} $
    \end{tabular}};
    \node[bias, rotate=90, minimum width=2cm] (r42) at (7, -3) {\scriptsize\begin{tabular}{c}
           $1\times3, C_1, d=2$ \\$\text{BN, ReLU} $
    \end{tabular}};
    \node[bias, rotate=90, minimum width=2cm] (r43) at (7, 0) {\scriptsize\begin{tabular}{c}
           $1\times3, C_1, d=3$ \\$\text{BN, ReLU} $
    \end{tabular}};
    \node[bias, rotate=90, minimum width=2cm] (r44) at (7, 3) {\scriptsize\begin{tabular}{c}
           $1\times3, C_1, d=4$ \\$\text{BN, ReLU} $
    \end{tabular}};
    
    \node[bias, rotate=90] (r5) at (8.2,0) {\scriptsize{Channel concatenation} };
    \node[h,rotate=90] (h3) at (8.7,0) {\scriptsize{Global Max Pool}};

    \node[bias, rotate=90, minimum width=1cm] (r6) at (9.5,0) {\scriptsize{Linear, $C_L$,  { ReLU}} };
    
    \node[bias, rotate=90, minimum width=2cm] (r7) at (10.2,0) {\scriptsize{Linear, 2} };
    \node[h,rotate=90,minimum width=2cm] (h4) at (10.7,0) {\scriptsize{Soft Max}};
    
    \node (y) at (11.5,0) {\small$y$};

    \draw[->] (x) -- (r1);
    \draw[] (r1) -- (h1);
    
    \draw[->] (h1) -- (r21);
    \draw[->] (h1) -- (r22);
    \draw[->] (h1) -- (r23);
    \draw[->] (h1) -- (r24);
    
    \draw [decorate,decoration={brace,amplitude=4pt}](3.5, 4.5) -- (6,4.5) node[midway, yshift=0.6cm] {\scriptsize$(M-1)$};
    \draw[->] (r21) -- (r3);
    \draw[->] (r22) -- (r3);
    \draw[->] (r23) -- (r3);
    \draw[->] (r24) -- (r3);
    \draw[] (r3) -- (h2);
    
    \draw[->] (h2) -- (r41);
    \draw[->] (h2) -- (r42);
    \draw[->] (h2) -- (r43);
    \draw[->] (h2) -- (r44);
    
    \draw[->] (r41) -- (r5);
    \draw[->] (r42) -- (r5);
    \draw[->] (r43) -- (r5);
    \draw[->] (r44) -- (r5);
    \draw[] (r5) -- (h3);
    
    \draw [decorate,decoration={brace,amplitude=4pt}](9, 2) -- (10,2) node[midway, yshift=0.6cm] {\scriptsize$2\times$};
    \draw[->] (h3) -- (r6);
    
    \draw[->] (r6) -- (r7);
    \draw[] (r7) -- (h4);
    
    \draw[->] (h4) -- (y);

\end{tikzpicture}